\documentclass[reprint,floatfix,aps,prx,superscriptaddress,longbibliography]{revtex4-2}

\usepackage{amsmath}
\usepackage{enumerate}
\usepackage{graphicx}
\usepackage{color}
\newcommand{\cred}{\color{red}}

\usepackage{cleveref}
\usepackage{mathrsfs}
\usepackage{xcolor}
\mathchardef\mhyphen="2D

\begin{document}

\title{Geometric control of universal hydrodynamic flow \\ in a two dimensional electron fluid}
\author{Ayd\i n Cem Keser}
\thanks{These authors contributed equally}
\author{Daisy Q. Wang}
\thanks{These authors contributed equally}
\affiliation{School of Physics, University of New South Wales, Sydney, NSW 2052, Australia}
\affiliation{Australian Research Council Centre of Excellence in Future Low-Energy Electronics Technologies, University of New South Wales, Sydney 2052, Australia}

\author{Oleh Klochan}
\affiliation{School of Science, University of New South Wales, Canberra ACT 2612, Australia}
%\affiliation{School of Physics, University of New South Wales, Sydney, NSW 2052, Australia}
\affiliation{Australian Research Council Centre of Excellence in Future Low-Energy Electronics Technologies, University of New South Wales, Sydney 2052, Australia}

\author{Derek Y. H. Ho}
\affiliation{Yale-NUS College, 16 College Avenue West, 138614, Singapore}

\author{Olga A. Tkachenko}
\affiliation{ Rzhanov Institute of Semiconductor Physics of SB RAS, Novosibirsk, 630090, Russia}

\author{Vitaly A. Tkachenko}
\affiliation{ Rzhanov Institute of Semiconductor Physics of SB RAS, Novosibirsk, 630090, Russia}
\affiliation{Novosibirsk  University, Novosibirsk, 630090, Russia}

\author{Dimitrie Culcer}
\affiliation{School of Physics, University of New South Wales, Sydney, NSW 2052, Australia}
\affiliation{Australian Research Council Centre of Excellence in Future Low-Energy Electronics Technologies, University of New South Wales, Sydney 2052, Australia}

\author{Shaffique Adam}
\affiliation{Yale-NUS College, 16 College Avenue West, 138614, Singapore}
\affiliation{Centre for Advanced 2D Materials and Graphene Research Centre, National University of Singapore, 6 Science Drive 2, 117546, Singapore}
\affiliation{Department of Physics, Faculty of Science
National University of Singapore, Science Drive 3,
Singapore 117551}

\author{Ian Farrer}
\altaffiliation[Present Address: ]{Department of Electronic and Electrical Engineering, The University of Sheffield, Mappin Street, Sheffield, S1 3JD, United Kingdom}  
\affiliation{Cavendish Laboratory, J. J. Thomson Avenue, Cambridge, CB3 0HE, United Kingdom}

\author{David A. Ritchie}
\affiliation{Cavendish Laboratory, J. J. Thomson Avenue, Cambridge, CB3 0HE, United Kingdom}

\author{Oleg P. Sushkov}
\affiliation{School of Physics, University of New South Wales, Sydney, NSW 2052, Australia}
\affiliation{Australian Research Council Centre of Excellence in Future Low-Energy Electronics Technologies, University of New South Wales, Sydney 2052, Australia}

\author{Alexander R. Hamilton}
\affiliation{School of Physics, University of New South Wales, Sydney, NSW 2052, Australia}
\affiliation{Australian Research Council Centre of Excellence in Future Low-Energy Electronics Technologies, University of New South Wales, Sydney 2052, Australia}
\date{\today}

%%%%%%%%%%%%%
%abstract%
%%%%%%%%%%%%%
\begin{abstract}

  Fluid dynamics is one of the cornerstones of modern physics and has recently found applications in the transport of  electrons in solids.
  In most solids electron transport is dominated by extrinsic factors, such as sample geometry and scattering from impurities.
%, and is essentially independent of the intrinsic properties of the electron system. 
However in the hydrodynamic regime Coulomb interactions transform the electron motion from independent particles to the collective motion of a viscous `electron fluid'. The fluid viscosity is an intrinsic property of the electron system, determined solely by the electron-electron interactions. Resolving the universal intrinsic viscosity is challenging, as it only affects the resistance through interactions with the sample boundaries, whose roughness is not only unknown but also varies from device to device. 
Here we eliminate all unknown parameters by fabricating samples with smooth sidewalls to achieve the perfect slip boundary condition, which has been elusive both in molecular fluids and electronic systems. We engineer the device geometry to create viscous dissipation and reveal the true intrinsic hydrodynamic properties of a 2D system. We observe a clear transition from ballistic to hydrodynamic electron motion, driven by both temperature and magnetic field. We directly measure the viscosity and electron-electron scattering lifetime (the Fermi quasiparticle lifetime) over a wide temperature range without fitting parameters, and show they have a strong dependence on electron density that cannot be explained by conventional theories based on the Random Phase Approximation.

\end{abstract}

%%%%%%%%%%%%%
\maketitle

\section{Introduction}
Fluid dynamics is one of cornerstones of modern physics and  technology, with wide ranging applications. Although a well-established subject (Bernoulli's law was formulated in 1738), it has important modern manifestations such as hydrodynamics of the quark-gluon plasma and of electrons in solids. While the dynamics of fluids are universal and depend only on the viscosity, the boundary conditions between the fluid and the containing solid play a crucial role. These boundary conditions are non universal, and depend on the details of the solid surface, the fluid, ambient conditions and the structure of complex boundary layers.  The precise nature of fluid boundary conditions at various interfaces is a long standing problem of great practical importance.

Hydrodynamic flow of electrons in solids occurs when extrinsic momentum-relaxing processes, such as electron-phonon and electron-impurity collisions, are much slower than intrinsic electron-electron scattering processes, which conserve the fluid's momentum~\cite{Gurzhi_1968}. Recent studies in graphene and other clean 2D systems have demonstrated viscous electron flow through Poiseuille flow~\cite{Sulpizzio_viscous, Ku20}, thermal and electrical transport effects~\cite{Bandurin, Nature_superballistic, Bandurin_onset, Gooth_thermal,Crossno, Moll, Gusev20, Govorov_jets}, and modifications to the Hall effect~\cite{Berdyugin_Hall,Gusev_Hall}. However the boundary problem remains unresolved: most studies have been performed in systems with diffusive boundaries~\cite{Molenkamp, Moll, Gusev18, Sulpizzio_viscous}. This introduces a system dependent unknown parameter, the `slip-length', which can vary with experimental conditions such as temperature and magnetic field, inhibiting quantitative analysis of experimental data. A perfectly smooth boundary condition would eliminate this unknown, allowing a direct measurement of the viscosity and hence the Fermi liquid quasiparticle lifetime (since viscosity depends only on electron-electron scattering). However the perfect slip boundary condition has been elusive and remains a mathematical idealization in the literature. 

% The search for hydrodynamic electron transport has largely relied on finding solids with sufficiently low disorder that electron-electron scattering dominates. Recent progress in fabricating ultra-clean semiconductor heterostructures and graphene samples has enabled the observation of viscous electron flow through vorticity of electrons~\cite{Bandurin, Nature_superballistic, Bandurin_onset}, Poiseuille flow~\cite{Sulpizzio_viscous, Ku20}, thermal or electrical transport effects~\cite{Gooth_thermal,Crossno, Moll, Gusev20, Govorov_jets}, and modifications to the Hall effect~\cite{Berdyugin_Hall,Gusev_Hall}. Though signatures of electron visocity have been captured, quantitatively extracting the quasiparticle lifetime is much more challenging. The fact that electron-electron interactions conserve the overall momentum makes it difficult to identify them in transport. They are primarily visible through the fluid's interplay with the system boundaries, where momentum relaxation occurs. 

\begin{figure*}[!ht]
\centering
	\includegraphics[width=\textwidth]{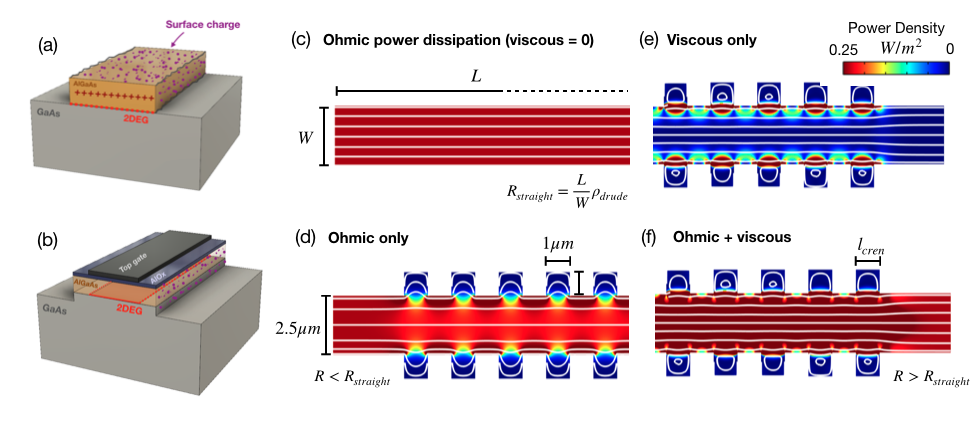}
    \caption{ \small (a) and (b) 3D schematics showing (a) A conventional modulation-doped GaAs/Al$_x$Ga$_{1-x}$As heterostructure. The conduction channel is patterned using chemical etching which causes rough sidewalls, while random surface charge on the sidewalls creates additional disorder. (b) An accumulation-mode GaAs/Al$_x$Ga$_{1-x}$As device. The channel is defined by a metal top-gate, and kept  away from etched sidewalls and surface charge. (c)-(f) Theoretical simulations (See sections I, IV and V Ref.~\onlinecite{Supplementary_material}) of the power dissipation density under the perfect slip boundary condition (width $W=2.5\:\mu m$, length $L=25\:\mu m$) in the linear response regime (Reynolds number much smaller than 1). % with $\rho_{Drude} = 40\:\Omega$ and $I = 200\:\text{nA}$.
    The colour scale represents the magnitude of the dissipation, and white lines show the electron flow streamlines. (c) Power dissipation density of a straight channel. The dissipation is purely ohmic as viscous contribution vanishes with perfect slip boundaries. (d) Ohmic power dissipation density of a crenellated channel (crenellation size $l_{\textrm{cren}}=1\:\mu\text{m}$) with zero viscosity, i.e. non-viscous electron transport. The ohmic resistance of the crenellated channel is smaller than the straight channel. (e) Viscous power dissipation density of the crenellated channel with a viscosity of $\:\nu=1.15\times10^{-2}\text{m}^{2}/\text{s}$. The viscous power dissipation density concentrates around the regions where the streamlines deform the most to form slow whirlpools in the crenellations. The viscous power dissipation vanishes near the boundaries due to the perfect slip boundary condition. (f) Total power dissipation density of the crenellated channel for a viscous electron flow summing up both ohmic and viscous contributions.}
	\label{fig:design}
\end{figure*}

The central idea of our work is to create devices with
perfectly smooth sidewalls. This eliminates unknown boundary effects, and constitutes the first realisation of universal viscous flow with the perfect slip condition. We first demonstrate perfect slip boundary conditions with no viscous dissipation in straight channels, and then controllably introduce viscous dissipation by carefully engineering the device geometry. The fluid flow is now determined solely by the geometry, hence bearing the name `universal hydrodynamic flow'.

% Experimental realisation of perfect slip condition regime is firstly a fundamental result in the general context of fluid dynamics.  Specifically for electron hydrodynamics, the universal hydrodynamic flow implies that we can controllably introduce viscous dissipation by carefully engineering obstacles to the flow. Since we do not need any fitting parameters to describe data, we unambiguously study and measure intrinsic properties of the electron fluid, i.e. the viscosity arising from electron-electron interactions.

Using this approach we observe a clear transition from ballistic to
hydrodynamic electron motion, driven by both temperature (which is expected)
and also by magnetic field (which is not). Moreover the absence of unknown boundary conditions allows \emph{quantitative} extraction of the viscosity and hence Fermi liquid quasiparticle lifetime over a wide temperature range, from $T\ll E_F$ to $T\sim E_F$. The experimental data reveal an unexpected and unexplained deviation of the electron-electron scattering length from existing theoretical models.

\section{Hydrodynamics in samples with smooth boundaries}
Describing the hydrodynamic flow of a fluid has two ingredients, (i) the dynamic Navier Stokes equation, (ii) the boundary condition at the fluid-solid interface. While the former is universal, the latter is not.

We begin by considering a straight channel with perfectly smooth  boundaries. No matter how strong the electron-electron interactions there is no viscous contribution in this straight channel, since the electron flow is uniform as shown in Figure~\ref{fig:design}(c). The resistance $R_{\textrm{straight}} = \rho_{Drude} L/W$, arises purely from phonons and impurity scattering. Viscous transport is introduced by modifying the device geometry with artificially engineered crenellations on the channel sidewalls, which causes non-uniform electron flow and thus viscous power dissipation as shown in Figure~\ref{fig:design}(d)-(f).
\subsection{Advantages of smooth boundaries}
The advantage of the smooth sidewalls with engineered structures lies in three aspects: 
\begin{enumerate}
    \item[(i)] All uncertainties associated with the slip length and boundary conditions are eliminated. 
    \item[(ii)] The electron transport regime can be unambiguously identified, simply by comparing the resistance of the straight channel and the crenellated channel. In transport measurements of samples with rough boundaries, it is hard to determine if the resistance $R(T)$ is due to viscous effects or scattering from extrinsic impurities and phonons. In our design when the transport is dominated by scattering with phonons and impurities, the wider crenellated channel will have a lower resistance than the straight channel $R_{\textrm{cren}}< R_{\textrm{straight}}$. This is because the crenellated channel is on average wider for the same length. However, in the hydrodynamic regime where electron-electron scattering dominates, the additional viscous contribution will increase the resistance of the crenellated channel so that $R_{\textrm{cren}}> R_{\textrm{straight}}$. 
    \item[(iii)]  Most importantly all experimental parameters associated with phonon and impurity scattering can be quantified through a direct comparison of the resistance of the straight and crenellated channels. Even when the electron transport is deep in the hydrodynamic regime, the resistance of the straight channel does not have any viscous contribution. This allows the electron-phonon and electron-impurity scattering processes to be fully characterised, so that the electron viscosity can be directly determined from the resistance of the crenellated channel without any unknown fitting parameters.
\end{enumerate}
%(i) (ii) (iii)

%In this report we present the measurement of the electron-electron scattering lifetime in 2D Fermi liquid via the viscosity of a hydrodynamic electron fluid. We utilise a device with no-stress side-walls in which both the non-viscous contributions and the viscous contribution to the electrical resistance can be experimentally controlled. We eliminate the uncertainty of extracting viscosity due to unknown parameters such as electron-phonon scattering rate, electron-impurity scattering rate and implicit boundary conditions. From the extracted electron-electron lifetime, we find a surprising density dependence than previous expected.
\subsection{Experimental realisation and verification of the perfect slip condition}
Experimentally the key challenge is how to make samples with perfect slip boundaries. In most conventional devices, such as graphene or the modulation-doped GaAs/AlGaAs heterostructure shown in Figure~\ref{fig:design}(a), chemical etching is required to pattern the channel of the 2D system, creating microscopically rough sidewalls. Moreover, random surface charge on the sidewalls of the channel creates additional disorder at the boundary. To avoid the uncontrolled roughness and disorder that cause rough boundary conditions we utilize an accumulation-mode GaAs/AlGaAs heterostructure depicted in Figure~\ref{fig:design}(b). There is no chemical doping in these accumulation-mode devices, and the  conduction channel is induced by applying a positive bias to the metallic top-gate. This ensures that the 2DEG is kept away from etched sidewalls and surface charge, providing a very smooth boundary.

%\subsection{The Device}
The device used in this study is divided into multiple segments containing both straight and crenellated channels, in which the resistance of each segment can be measured independently. Figure~\ref{fig:device}(a) shows a $2.5\:\mu\text{m}$ wide by $25\:\mu\text{m}$ long straight channel, adjacent to a crenellated channel of the same length, with a minimum width varying between $2.5\:\mu\text{m}$ and $4.5\mu\text{m}$, due to $1\:\mu\text{m}$ by $1\:\mu\text{m}$ crenellations.
Figure~\ref{fig:device}(b) shows the calculated characteristic length scales for this device~\cite{Supplementary_material}. For $10\:\text{K} \lesssim T\lesssim40\:\text{K}$, the electron-electron scattering length $l_{ee}$ is the shortest length scale in the system, so that hydrodynamic effects are significant. At very low temperatures $T \lesssim 10\:\text{K}$, $l_{ee}$ exceeds the characteristic length scale of the device $l_{\textrm{cren}}\sim1\mu\text{m}$, and electron transport is ballistic.

%\subsection{Experimental verification of the perfect slip condition}
We verify the smooth boundary condition in our experiment by comparing the resistance of the two straight channels with different widths: $5\:\mu\text{m}$ and $2.5\:\mu\text{m}$ (both have the same length to width ratio of $10$). The two straight channels have almost identical resistance across the whole temperature range for all three densities, shown by the red circles and grey squares in Figure~\ref{fig:device}(c)-(e). This proves that the perfect slip boundary condition is satisfied in our device,
since the resistance of channels with rough boundaries has a strong width dependence in both in the hydrodynamic~\cite{Gurzhi_1968} and ballistic~\cite{beenakker} regimes. 
\section{Disentangling diffusive and viscous transport}

\subsection{Straight Channel}
One of the key challenges in quantitative extraction of the viscosity of the electron fluid is how to disentangle viscous and non-viscous contributions to the resistance. A unique advantage of the perfect slip boundaries is that the resistivity of the straight channels has no ballistic back scattering or viscous components, hence serves as an absolute reference  from  which  all  momentum-relaxing  contributions  can  be  measured: (i) From the resistance at base temperature $T=0.25\:\text{K}$, we calculate the momentum relaxation length due to impurity scattering $l_{imp}=v_F\tau_{imp}\approx10\mu \text{m}$, corresponding to a mobility on the order of $10^6\: \text{cm}^2/\text{Vs}$ (see sections I and II of the supplement~\cite{Supplementary_material}). %(Supplement secs.~I-II).
(ii) The linear increase of $R_{\textrm{straight}}(T)$ with temperature is due to phonon scattering. We extract the phonon scattering time $\tau_{ph} = A^{\tau}_{ph}/T$ from the slope $d R_{\textrm{straight}}(T)/d T$.  This gives a phonon coupling constant of $A^{\tau}_{ph} = 1.5\:\text{ns}\cdot\text{K}$, consistent with previous studies of electrons in GaAs~\cite{Gusev18,Shi} (sec. III of the supplement~\cite{Supplementary_material}). % (See Supplement sec.~III).

\begin{figure*}[!ht]
\centering
	\includegraphics[width=\textwidth]{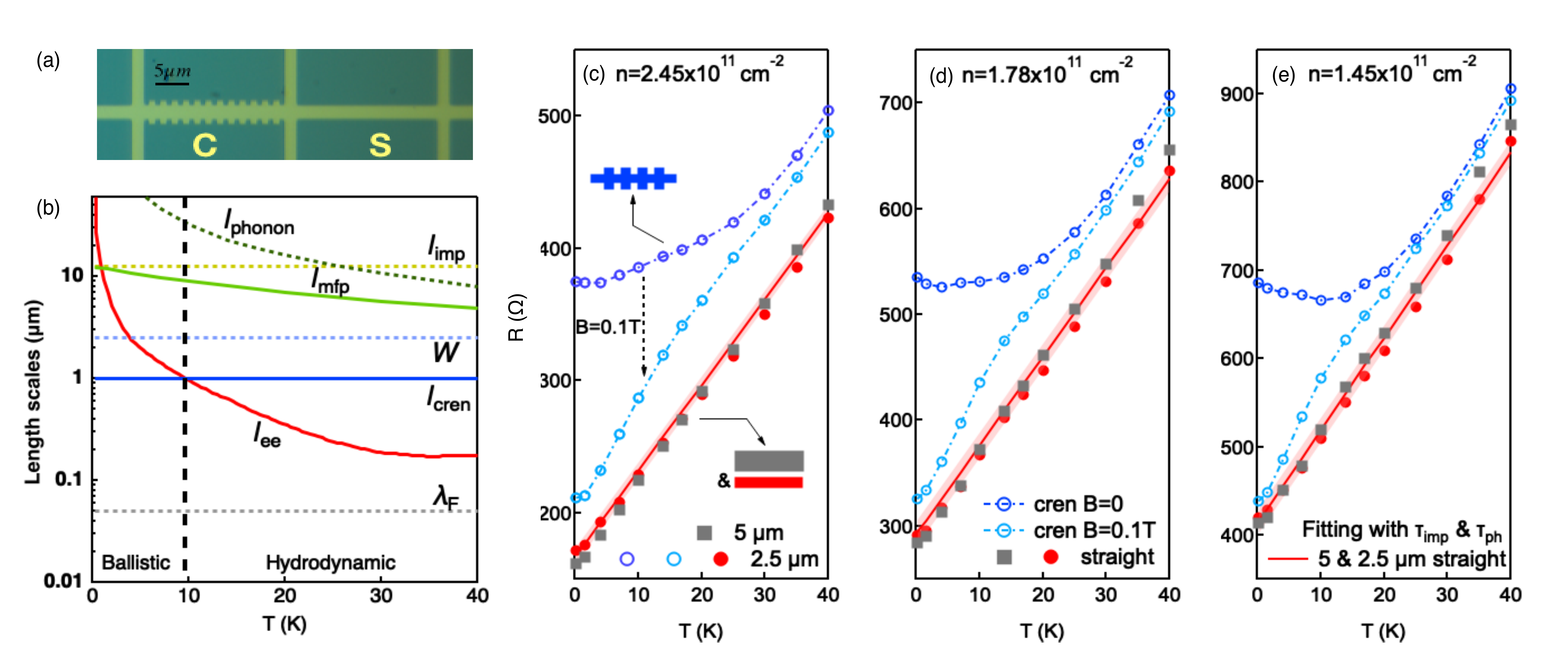}
	\caption{\small (a) Image of the accumulation mode GaAs heterostructure device containing a high mobility 2DEG. The gate-defined channel contains two sections: a $2.5\:\mu \mathrm{m}$ wide and $25\:\mu \mathrm{m}$ long straight channel labelled ``S", and a crenellated channel ``C" of the same dimensions with crenellations of $l_{\textrm{cren}}=1\:\mu \mathrm{m}$. (b) Key length scales and transport regimes in the system. $l_{mfp}$ is the momentum-relaxing mean free path due to electron-phonon ($l_{\textrm{phonon}}$) and electron-impurity ($l_{imp}$) scattering. $W$ is the width of the channel and $l_{\textrm{cren}}$ is the size and spacing of the square crenellations. $\lambda_F$ is the Fermi wavelength. $l_{ee}$ is the theoretically calculated %(Supplement sections VI-VII) 
	electron-electron scattering length $l_{ee}$. When $T < 10\:\text{K}$, $l_{\textrm{cren}}$ is the smallest length scale and the system is in the ballistic transport regime. When $l_{ee}$ falls below $l_{\textrm{cren}}$ at $T>10\:\mathrm{K}$, the system enters the hydrodynamic transport regime. 
	(c)-(d) Experimentally measured resistance of both straight and crenellated channels 
	$R_{\textrm{straight}}$, $R_{\textrm{cren}}$ as a function of temperature for three electron densities 
	$n = 2.45,1.78,1.45\:\times 10^{11}\:\text{cm}^{-2}$, respectively. 
	The resistance of two straight channels with different widths $W = 2.5,5\:\mu\text{m}$ but the same 
	length to width ratio are shown by the red solid circles and grey squares, and are indistinguishable. The solid red lines are fits to $R_{\textrm{straight}}$ including both 
	electron-impurity and electron-phonon scattering with the shaded area presenting the 
	uncertainty. The resistance of the crenellated channel``C" is shown both at $B=0$ (dark blue empty circles) and with a small out of plane 
	magnetic field $B=0.1\mathrm{T}$ (light blue empty circles).}
	\label{fig:device}
\end{figure*}

\begin{figure*}
\centering
	\includegraphics[width=\textwidth]{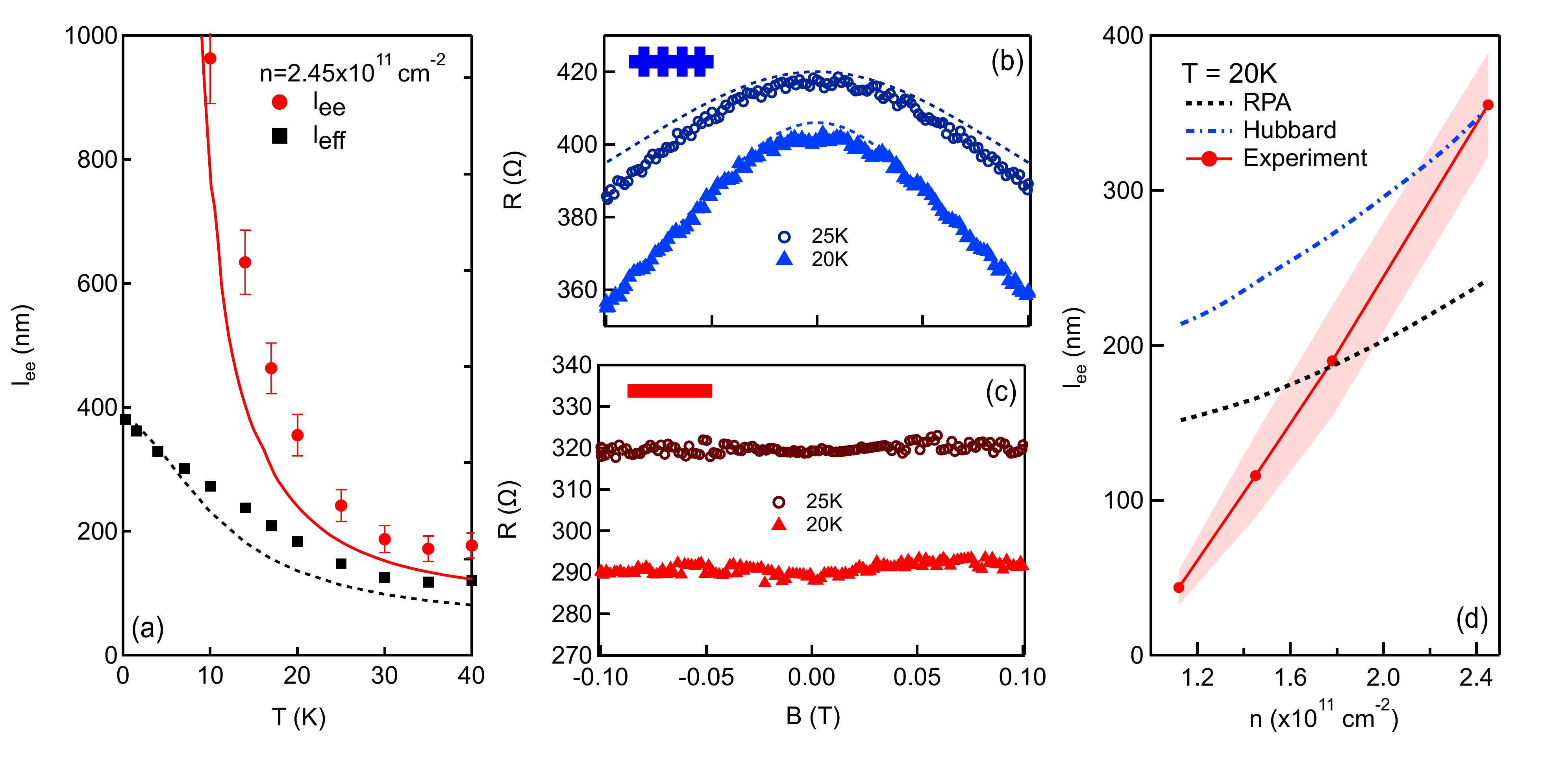}
	\caption{\small (a) Electron-electron scattering length $l_{ee}$ and the viscous effective scattering length $l_{\textrm{eff}}$ as a function of temperature for an electron density of $n=2.45\times10^{11}\text{cm}^{-2}$. The solid red line is the parameter free theoretical calculation of $l_{ee}$ for a Fermi liquid using RPA and Boltzmann theory (See section VI of the supplement~\cite{Supplementary_material}). The dashed black is the calculated $l_{\textrm{eff}}$ using Eq.~\ref{parallel}. The black squares and red filled circles are the values of $l_{\textrm{eff}}$  and $l_{ee}$ extracted from the measured resistance of the crenellated channel. The error bars are due to the uncertainty in the ohmic component of resistance %(Supplement secs.~I and II).  
	  (b,c) High-temperature magnetoresistance of the crenellated and straight channels for $n=2.45\times10^{11}\text{cm}^{-2}$.
The straight channel,panel (c), shows no magnetoresistance as expected for smooth sidewalls, whereas the crenellated channel,  panel (b),
shows a strong negative magnetoresistance due to viscous effects. The dashed lines in (b) are theoretical magneto-resistance curves, calculated using Eq.~\eqref{magneto} and the $l_{ee}$ measured in panel (a). %Magneto-resistances of (d) the crenellated channel and (e) the straight channel at the same electron density but low temperatures $T=0.25\mathrm{K}$ and $T=4\mathrm{K}$, respectively. 
	(d) Density dependence of $l_{ee}$ at $T=20\:\mathrm{K}$. Values extracted from the experiment (symbols, with error margin shown by the shading) are a stronger function of the electron density  than predicted theoretically from the RPA (black dashed line) and Hubbard approximation (blue dash-dotted line). 
	%The interaction parameter $r_s = 1.17\mhyphen1.72$. (See Supplement sections~VI and VII for further details.)
	}
	\label{fig:lee}
\end{figure*}

\subsection{Crenellated Channel}
The resistance of the crenellated channel with a width of $2.5\:\mu\text{m}$ and crenellations of $l_{\textrm{cren}}=1\:\mu\text{m}$ is shown as the dark blue circles in Figure~\ref{fig:device}(c)-(e). The crenellated channel resistance is always higher than that of the straight channels, despite having the same length and minimum width as the $2.5\mu\text{m}$ wide straight channel. % (See Supplement sec.~IV). 
At high temperatures $R_{\textrm{cren}}(T)$ exhibits a close to linear dependence on $T$ with the same slope as $R_{\textrm{straight}}(T)$. As the temperature is lowered, $R_{\textrm{cren}}(T)$ starts to deviate from the linear dependence and rapidly increases when $T<10K$, which is particularly visible at low electron densities. This non-monotonic behaviour of $R_{\textrm{cren}}(T)$ at low temperatures is caused by quasiballistic effects at low temperatures~\cite{Alekseev-Semina}, which hide the viscous behaviour~\cite{Bandurin_onset, Gusev18, Nature_superballistic, Sulpizzio_viscous}. 
Disentangling the viscous, diffusive and ballistic effects can be very complicated both theoretically and experimentally when the boundary condition is unknown~\cite{Sulpizzio_viscous,Nature_superballistic}, but is simple for smooth boundaries. In this limit ballistic effects can be easily excluded by applying a small perpendicular magnetic field of $B = 0.1\:\text{T}$.
The magnetic field suppresses ballistic back-scattering, since the cyclotron radius $r_c \approx 800\:\text{nm}$ is shorter than the smallest feature size of the channel, while the viscous friction stays robust, given that $r_c$ is about twice larger than the typical electron-electron scattering length $l_{ee}$. The light blue circles in Figure~\ref{fig:device}(c)-(e) show the resistance of the crenellated channel at $B=0.1\:\text{T}$. The low temperature resistance is reduced by the suppression of ballistic effects, but is still larger than the straight channel. This confirms that there is a significant viscous contribution to the resistance of the crenellated channel over a wide temperature range.
At higher temperatures the resistance at $B=0.1\:\text{T}$ approaches the $B=0\:\text{T}$ resistance, as ballistic contributions decline. There is a clear change of slope in the $B=0.1\:\text{T}$ data at $T\approx10\:\text{K}$ which marks a soft transition from ballistic to hydrodynamic transport regimes~\cite{Alekseev-Semina}. This is consistent with the crossover temperature expected from the length scales of the system shown in Figure~\ref{fig:device}(b). 

\section{Extraction of viscosity \& $l_{ee}$}
To quantitatively extract the viscosity $\nu$ from the experimental data we solve the Navier-Stokes equations with perfect slip (no-stress) boundary conditions ~\cite{Pellegrino,Levitov_non-local,Supplementary_material}: 
%\begin{subequations}
    \begin{align}
    \mathbf{v}/\tau_{mfp} + \mathbf{v} \cdot \nabla \mathbf{v} - \nu \nabla^2 \mathbf{v} &= -\nabla \Phi/m^*,\nonumber \\
    \nabla \cdot \mathbf{v} &= 0,
   % \label{incompressible}
   \label{NSE}
    \end{align}
Here $\mathbf{v}$ is the velocity field, $\tau_{mfp}$ is the mean free time due to phonons and disorder, $m^*$ is the effective mass of electron, $\nu$ is the kinematic viscosity and $\Phi$ is the electrochemical potential. The mean free time is extracted directly from the measured resistance of the straight channels, $R_{\textrm{straight}} = m^* L/(n W e^2 \tau_{mfp})$. Hence the viscosity $\nu$ is the only unknown parameter. We chose a particular value of $\nu$,  and numerically solve the Navier-Stokes equations (see secs.~IV and V of the supplement~\cite{Supplementary_material}).  The numerical solution of equation~\ref{NSE} for a given $\nu$ gives the velocity field, from which we calculate the dissipation and compare it with the measured resistance. We repeat this procedure until we find the value of $\nu$ that reproduces the experimental resistance of  the crenellated segment. This is the ``experimental value'' of $\nu$. For an infinitely large device the relation to the electron-electron scattering length $l_{ee}$ is $\nu=\frac{1}{4}v_Fl_{ee}$ (see secs.~VI and VIIA of the supplement~\cite{Supplementary_material}). However, for a real device we need to account for ballistic effects.
  
\subsection{Ballistic effects}
At low temperatures $l_{ee}$ approaches the critical sample dimensions (the crenellation length $l_{\textrm{cren}}$ in our case), and this finite size effect causes an apparent saturation of $l_{ee}$ as $T\to0$. Following Ref.~\onlinecite{Nature_superballistic}, we capture both the finite size and ballistic effects by introducing an effective viscous mean free path $l_{\textrm{eff}}$
\begin{equation}
\label{parallel}
    \nu = \frac{1}{4}v_F l_{\textrm{eff}},\quad \frac{1}{l_{\textrm{eff}}(T)} = \frac{1}{l_{ee}(T)} + \frac{1}{l_{\textrm{eff}}(T=0)}.
\end{equation}
Here $l_{\textrm{eff}}$ is completely independent of scattering of electrons from phonons and impurities, and captures the ballistic and viscous effects. The cut-off at the ballistic limit $l_{\textrm{eff}}(T = 0) \approx l_{\textrm{cren}}/2.5$  is not a fitting parameter,  but is extracted  from  the  base  temperature $T=0.25\:\text{K}$ measurement which is dominated by ballistic effects (see sec.~VIIA of the supplement~\cite{Supplementary_material}).

We plot the extracted electron-electron scattering lengths with symbols in Figure~\ref{fig:lee}(a) as a function of temperature for an electron density of $n=2.45\times10^{11}\:\text{cm}^{-2}$. (See Figure~S4 of the supplement~\cite{Supplementary_material} for other densities.) Both $l_{\textrm{eff}}$ and $l_{ee}$ increase with decreasing $T$, with $l_{\textrm{eff}}$ saturating as it approaches half the crenellation length scale. In contrast, $l_{ee}$, an intrinsic property of the electron liquid, diverges as $T\rightarrow0$. 

It is well known that at very low temperatures the electron-electron
  scattering length scales $l_{ee}\propto 1/T^2$. However, this approximation is valid only at $T\ll 0.1 E_F$ (see Fig.S3b of the supplement.~\cite{Supplementary_material}), and most data in the
 literature, including ours, are obtained at $T> 0.1 E_F$. Therefore we go beyond
the low temperature approximation, as described in Sec.~VI and VII of the supplement~\cite{Supplementary_material}. 
The solid and dashed lines show our theoretical calculations of $l_{ee}$ and $l_{\textrm{eff}}$ using the random phase approximation (RPA) with $l_{\textrm{eff}}(T=0)$ as the only parameter taken from experiment. The parameter-free calculation of $l_{ee}$ is in remarkably good agreement with the experiment, given the lack of any fitting parameters. 

\section{Independent verification of $l_{ee}$ through magnetotransport}
To independently check the values of the extracted $l_{ee}$ we show in Figure~\ref{fig:lee}(b)-(c) low-field magneto-resistance measurements at the same carrier density. The crenellated channel exhibits a parabolic negative magneto-resistance, shown in Figure~\ref{fig:lee}(b), due to the magnetic suppression of the viscosity~\cite{Alekseev,hydro-hall, Steinberg}
\begin{equation}
    \nu(B) = \nu(0) \frac{B_*^2}{B_*^2 + B^2},\quad 
    B_* = \frac{p_F}{2 |e| l_{ee}}.
    \label{magneto}
\end{equation}
The characteristic magnetic field $B_*$ thus provides a direct measurement of the zero-field $l_{ee}$, without solving the Navier Stokes equations. Using the second part of Eq.~\eqref{magneto} and the experimentally determined values of $l_{ee}$ from Figure~\ref{fig:lee}(a) we find $B_* \approx 105, \:152 \:\text{mT}$ for $T=20,\:25\:\text{K}$, respectively. The magneto-resistance calculated from Eq.~\eqref{magneto} is shown by dashed lines in Figure~\ref{fig:lee}(b), and is in excellent agreement with the measurement. In contrast to the crenellated channels, the straight channels have no noticeable dependence on $B$ over the same field range, as shown in Figure~\ref{fig:lee}(c)). This confirms the absence of viscous effects in the straight channels, i.e. the boundaries are smooth. 
%At low temperatures, $T = 0.25\text{---}4\:\text{K}$, the crenellated segment $R_{\textrm{cren}}(B)$ have a distinct magneto-transport signature due to the ballistic size effects as seen by the double hump structure in Figure~\ref{fig:lee}(e). Around $B = 18\: \text{mT}$ the cyclotron radius $r_c$ is twice the width of the sample and the ballistic back-scattering is maximized, consistent with earlier studies~\cite{boundary_magneto}. However, beyond the maximum point, ballistic back-scattering is quickly eliminated by the magnetic field, leading to a characteristic peak in $R(B)$ when $r_c\sim2W$. 

\section{Discussion and Conclusions}

Comparing the $l_{ee}$ calculated using the parameter free theory with the measured values in Figure~\ref{fig:lee}(a), it is notable that the experimental values are consistently higher than theoretical predictions. 
This deviation is even more pronounced when examining the variation of $l_{ee}$ (and hence the electron quasiparticle lifetime) with carrier density. The $l_{ee}$ calculated in the RPA shows only a weak dependence on $n$, in contrast to the strong dependence measured experimentally in Figure~\ref{fig:lee}(d). These results suggest that $\tau_{ee}$ has a significant density dependence that is not captured in the RPA. This is especially surprising given the small interaction parameter $r_s \sim 1.1\mhyphen1.5$ of the electron system (See for example Ref.~\cite{Reza} and references therein). 
We have checked that this discrepancy is not an effect of the methods used to solve the Navier Stokes equations, nor can it be explained by including density dependent screening effects in the 2DEG. 
As shown in section VIIC of the supplement \onlinecite{Supplementary_material}, these have a small effect on $l_{ee}$, but do not influence the density dependence of $l_{ee}$. 
Furthermore we have gone beyond the RPA, using the Hubbard approximation (Supplement sec. VI and VII \cite{Supplementary_material}), but the calculated density dependence (blue dash-dotted line in Figure~\ref{fig:lee}(d)) remains  inconsistent with experiment. 
This suggests that correlation effects beyond the Hubbard approximation are significant even at relatively low $r_s$. 

In conclusion, we have created 2D electron channels with perfect slip boundaries, thereby eliminated unknown parameters related to boundary scattering. This method makes it possible to separate extrinsic (phonon and disorder) scattering effects from the intrinsic viscous effects due to electron-electron scattering. From the viscous resistance we directly extract the electron-electron scattering length. The techniques and analysis introduced here open a new route to probing the finite temperature quasiparticle lifetime of two-dimensional Fermi liquids over a wide temperature range.

\begin{acknowledgements}
%\section{Acknowledgements}
We acknowledge important discussions with Andre Almeida, Reza Asgari, Gennady V. Stupakov, Joe Wolfe and Igor Zutic.
This work was supported by the Australian Research Council Centre of Excellence in Future Low-Energy Electronics Technologies (CE170100039). D. A. Ritchie acknowledges support from the Engineering and Physical Sciences Research Council, United Kingdom. Devices were made at the NSW node of the Australian National Fabrication Facility. 
\end{acknowledgements}
{\cred}
\emph{Note added: After initial submission of this work we became aware of a study of viscous behaviour in a GaAs 2DEG using non-local measurements~\cite{Gupta}. The $l_{ee}$ differ from those obtained here; a comparison of the two approaches is given in the supplement~\cite{Supplementary_material}.
}
%% Put the bibliography here, most people will use BiBTeX in
%% which case the environment below should be replaced with
%% the \bibliography{} command.

%{\large{\bf Author contributions}}

%D.Q.W. fabricated the devices. D.Q.W. and O.K. performed the experiments. I.F. and D.A.R. grew the heterostructure. A.C.K., O.P.S, D.Y.H.H. and S.A. performed the theoretical calculations. O.A.T and V.A.T. performed finite element calculations of the electrostatic screening of the top gate. O.P.S. and A.R.H. conceived the project. A.C.K., D.Q.W., O.P.S., A.R.H., O.K., S.A. and D.C. analysed the data and prepared the manuscript with input from all authors.

%{\large{\bf Competing financial interests}}

%The authors declare no competing financial interests.

\bibliographystyle{apsrev4-2}
\bibliography{electron_fluid_refs}

\end{document}